\documentclass[aps,prl,reprint]{revtex4-2}

\usepackage{float}
\usepackage{amsmath,amssymb,amsfonts}
\usepackage{graphicx}
\usepackage{bm}
\usepackage{lineno}
\makeatletter
\renewcommand{\paragraph}[1]{%
  \par\vspace{0.6em}%
  \noindent\textit{#1 ---}
}
\makeatother
\begin{document}


\title{Geometric Curvature Governs Work in Open Quantum Steady States}

\author{Eric R. Bittner}
\affiliation{Department of Physics, University of Houston, Houston, TX, USA}

\date{\today}
\begin{abstract}
We show that quasistatic work in open quantum steady states is governed by an emergent geometric curvature in control-parameter space. This curvature arises from the interplay of coherent dynamics and dissipation and defines a work one-form whose flux determines the work produced in cyclic processes. Unlike classical thermodynamics, where work depends only on the area enclosed by a cycle, the geometric structure is spatially inhomogeneous: cycles enclosing comparable areas can yield different work depending on their location in parameter space. The curvature vanishes under strong dephasing, identifying steady-state coherence as a necessary condition for nontrivial geometric response. Reversing the cycle orientation reverses the sign of the work, confirming its geometric origin. We illustrate these results for a driven dissipative two-level system and show that the magnitude and sign of the work are set by the spatial structure of the curvature rather than by coherence alone. These results establish curvature as the organizing principle of thermodynamic response in open quantum systems, providing a geometric framework for driven light--matter systems in cavity quantum electrodynamics.
\end{abstract}

\maketitle

\paragraph{Introduction}
Classical thermodynamics admits a geometric formulation in which work is associated with areas enclosed by cycles in state space, $W = \oint P\,dV$. More generally, equilibrium thermodynamics can be expressed in terms of intrinsic geometric structures—metric, contact, and symplectic—in which equations of state define submanifolds of a higher-dimensional phase space.\cite{Weinhold1975,Ruppeiner1995,Grmela2014,vanDerSchaft2018} In this setting, thermodynamic response is encoded in geometric objects defined over the space of equilibrium states.

Stochastic thermodynamics extends these ideas to nonequilibrium systems by assigning thermodynamic quantities such as work, heat, and entropy production to individual trajectories.\cite{Seifert2012} These developments establish a consistent framework for energetics in driven systems and nonequilibrium steady states. However, existing approaches either describe geometric structure in thermodynamic state space or define work through stochastic trajectories, and do not identify a geometric object in control-parameter space that governs quasistatic work. In particular, no curvature has been identified whose flux determines work in driven dissipative quantum systems.

Open quantum systems provide a natural setting in which to address this question. Their steady states arise from the interplay of coherent dynamics and dissipation, typically described by Lindblad evolution.\cite{Lindblad1976,BreuerPetruccione} In driven light--matter systems, particularly in cavity quantum electrodynamics, coherent control and engineered dissipation are intrinsically intertwined, producing steady states that are generically not diagonal in the instantaneous energy basis. This raises a central question: can nonequilibrium steady states support a geometric thermodynamic description in parameter space?

Recent work has shown that geometric properties of quantum states, including the quantum geometric tensor and its finite-temperature generalizations, can be expressed in terms of dynamical correlation functions and accessed through linear response via fluctuation–dissipation relations.\cite{Ji2025,Carollo2018} In this framework, geometric quantities such as the quantum metric and Uhlmann curvature emerge as experimentally measurable response functions. Related developments in stochastic and quantum thermodynamics have emphasized the role of trajectory-dependent quantities and response functions in characterizing nonequilibrium processes.\cite{Crooks1999,Seifert2012,Esposito2009} However, these approaches characterize the geometry of states rather than that of thermodynamic processes.

Here we show that quasistatic work in open quantum steady states is governed by an emergent curvature defined over control-parameter space, arising from the non-integrable structure of the system’s steady-state response.\cite{Bittner2026_quantumGeom} This curvature originates from the parameter dependence of the steady-state density matrix and reflects the competition between coherent driving and dissipation, rather than any underlying equilibrium equation of state. As a result, work becomes intrinsically geometric: it depends not only on the area enclosed by a cycle, but also on its location in parameter space. The curvature vanishes under strong dephasing, identifying steady-state coherence as a necessary condition for nontrivial geometric response. Quasistatic work is therefore given by the flux of this curvature, establishing it as the organizing principle of thermodynamic response in driven dissipative quantum systems.

\paragraph{Geometric Work in Parameter Space}
We consider an open quantum system described by a Hamiltonian $H(\lambda)$ that depends on a set of control parameters $\lambda = \{\lambda_i\}$. Under slow (quasistatic) variation of $\lambda$, the system remains close to its instantaneous steady state $\rho_{\rm ss}(\lambda)$. The work differential is
\begin{equation}
\delta W = \mathrm{Tr}\!\left(\rho_{\rm ss} \, dH\right),\label{eq:deltaW}
\end{equation}
which defines a work one-form in parameter space,
\begin{equation}
\mathcal{A} = \sum_i A_i \, d\lambda_i,
\qquad
A_i = \mathrm{Tr}\!\left(\rho_{\rm ss} \, \partial_{\lambda_i} H\right).
\end{equation}
This expression shows that work arises from how the steady-state expectation value of the Hamiltonian changes under parameter variation. In contrast to equilibrium systems, where work can be expressed in terms of thermodynamic potentials, here it is determined directly by the response of the steady state to changes in the control parameters. We adopt the convention that $W$ denotes work done on the system, so that positive $W$ corresponds to energy input under parameter variation, consistent with Eq.~\ref{eq:deltaW}.

The work performed over a closed cycle $\gamma$ is
\begin{equation}
W_{\rm cyc} = \oint_{\mathcal{C}} \mathcal{A},
\end{equation}
where $\mathcal{C}$ denotes a closed cycle in parameter space, with $\mathcal{C}^{-1}$ its reversal.
Using Stokes' theorem, this can be expressed in terms of a curvature two-form,
\begin{equation}
W_{\rm cyc} = \iint_\Sigma \mathcal{F},
\qquad
\mathcal{F} = \frac{1}{2} \sum_{i,j} \mathcal{F}_{ij} \, d\lambda_i \wedge d\lambda_j, \label{eq:4}
\end{equation}
where
\begin{equation}
\mathcal{F}_{ij} = \partial_{\lambda_i} A_j - \partial_{\lambda_j} A_i. \label{eq:5}
\end{equation}
The curvature $\mathcal{F}_{ij}$ quantifies the noncommutativity of parameter variations: it measures the extent to which the work accumulated by varying $\lambda_i$ followed by $\lambda_j$ differs from that obtained by reversing the order. A nonzero curvature therefore signals intrinsically geometric, path-dependent work.

Physically, this noncommutativity reflects the non-integrable response of the steady state to parameter changes: successive variations probe distinct coherence structures of the steady-state manifold, producing path-dependent work.
Unlike equilibrium thermodynamics, where quasistatic work is path-independent, curvature implies that even quasistatic processes in open quantum systems are inherently path-dependent.

In two-dimensional parameter spaces, $\lambda = (\lambda_1,\lambda_2)$, the curvature reduces to the scalar curl of the vector potential $\mathbf{A} = (A_1,A_2)$,
\begin{equation}
\mathcal{F}_{12} = (\nabla \times \mathbf{A})_z, \label{eq:6}
\end{equation}
so that the work is given by the flux of this effective field through the enclosed cycle. In this form, quasistatic work is determined by the flux of an emergent geometric field in parameter space.

This construction is not restricted to the present model, but applies to any open quantum system admitting a steady state $\rho_{\rm ss}(\lambda)$ that depends smoothly on control parameters. In this setting, the work one-form $\mathcal{A}$ defines a connection over parameter space, and the curvature $\mathcal{F}$ encodes the intrinsic geometric response of the steady-state manifold. This structure is independent of any equilibrium equation of state and instead reflects the dynamical interplay between coherent evolution and dissipation, providing a general framework for quasistatic thermodynamic response in nonequilibrium quantum systems.

Thus, quasistatic work in an open quantum steady state is governed by a curvature defined over control-parameter space. This curvature is the nonequilibrium analogue of geometric structure in classical thermodynamics, but here emerges dynamically from the competition between coherent driving and dissipation. Crucially, steady-state coherence is not merely a quantitative resource: it is the mechanism that generates curvature and thereby enables geometric work.

The structure of Eqs.~\ref{eq:4}--\ref{eq:6} is formally analogous to a Berry curvature defined over parameter space, but constructed from a steady-state density matrix rather than adiabatic eigenstates.  However, in contrast to geometric phases in closed quantum systems, the curvature identified here is defined by the steady-state density matrix of an open system and encodes both dissipative and coherent dynamics. It therefore defines a genuinely nonequilibrium geometric structure. The resulting curvature is strongly localized, defining preferred regions of parameter space where geometric response is concentrated.

\begin{figure}[t]
\includegraphics[width=\columnwidth]{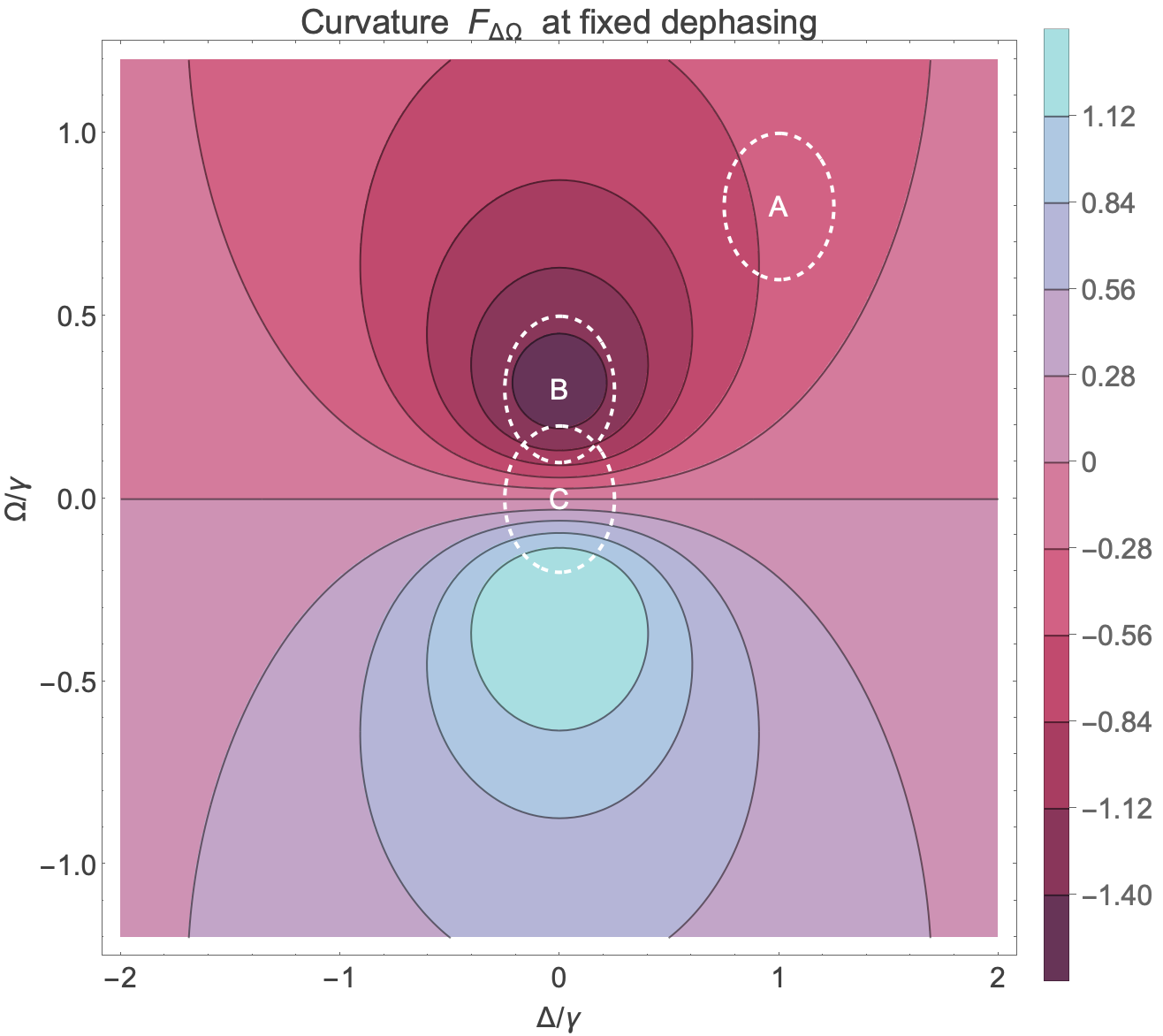}
\caption{
Curvature $\mathcal{F}_{\Delta\Omega}$ in the $(\Delta,\Omega)$ plane. Two representative cycles are shown: loop A in a weak-curvature region and loop B in a high-curvature region. The curvature is strongly localized near resonance.
}
\label{fig1}
\end{figure}

\paragraph{Model and Steady State}
To illustrate this construction, we consider a driven two-level system with control parameters $\lambda = (\Delta,\Omega)$ and Hamiltonian
\begin{equation}
H(\Delta,\Omega) = \frac{\Delta}{2}\sigma_z + \Omega \sigma_x.
\end{equation}
The system is coupled to a Markovian environment,
\begin{equation}
\dot{\rho} = -i[H,\rho] + \gamma \mathcal{D}_{\sigma_-}[\rho] + \frac{\gamma_\phi}{2}\mathcal{D}_{\sigma_z}[\rho],
\end{equation}
with relaxation rate $\gamma$ and pure dephasing rate $\gamma_\phi$.

In Bloch representation,
\begin{equation}
\rho = \frac{1}{2}\left(I + x \sigma_x + y \sigma_y + z \sigma_z\right),
\end{equation}
the steady-state components are
\begin{align}
x_{\rm ss} &= -\frac{2\gamma \Omega \Delta}{D}, \\
y_{\rm ss} &= \frac{2\gamma \Omega \Gamma_2}{D}, \\
z_{\rm ss} &= -\frac{\gamma(\Delta^2 + \Gamma_2^2)}{D},
\end{align}
where
\begin{equation}
\Gamma_2 = \frac{\gamma}{2} + \gamma_\phi,
\qquad
D = 4\Omega^2 \Gamma_2 + \gamma(\Delta^2 + \Gamma_2^2).
\end{equation}

Because the Hamiltonian eigenbasis is generically misaligned with the dissipative basis, the steady state acquires finite coherence and is not diagonal in the energy basis. This misalignment is the microscopic origin of the geometric structure: it enforces a nontrivial parameter dependence of the steady state, producing a finite curvature. Substituting these expressions into the general definition of the work one-form yields
\begin{equation}
\mathcal{A} = A_\Delta d\Delta + A_\Omega d\Omega,
\end{equation}
with
\begin{align}
A_\Delta &= \frac{1}{2} z_{\rm ss}, \\
A_\Omega &= x_{\rm ss},
\end{align}
and the corresponding curvature
\begin{equation}
\mathcal{F}_{\Delta\Omega} = \partial_\Delta A_\Omega - \partial_\Omega A_\Delta.
\end{equation}

Evaluating explicitly,
\begin{equation}
\mathcal{F}_{\Delta\Omega}
=
-\frac{2\Omega\gamma\left(2\gamma_\phi \Delta^2 + \Gamma_2(2\Gamma_2^2 + \Gamma_2\gamma + 4\Omega^2)\right)}{D^2}.
\end{equation}

Figure~\ref{fig1} shows $\mathcal{F}_{\Delta\Omega}$ in the $(\Delta,\Omega)$ plane. The curvature is strongly nonuniform and exhibits a pronounced maximum near resonance, where coherent driving competes most strongly with dissipation. Two representative cycles are indicated: loop A in a weak-curvature region and loop B in a high-curvature region. This implies that the geometric work depends on the local distribution of curvature over the enclosed region rather than the area alone.

\paragraph{Coherence as a Necessary Resource}
The steady-state coherence in the dissipative basis is
\begin{equation}
C_{\rm bare} = \frac{1}{2}\sqrt{x_{\rm ss}^2 + y_{\rm ss}^2}.
\end{equation}

In the strong dephasing limit $\gamma_\phi \to \infty$, the coherences vanish, $\{x_{\rm ss},y_{\rm ss}\} \to 0$, and the steady state becomes diagonal in the dissipative basis. In this limit, both the curvature and the associated cycle work vanish,
\begin{equation}
\mathcal{F}_{\Delta\Omega} \to 0,
\qquad
W_{\rm cyc} \to 0.
\end{equation}

Thus, steady-state coherence is a necessary condition for nonzero geometric work. However, coherence alone does not determine the magnitude of the response. Instead, the work is governed by the curvature of the steady-state manifold, which encodes how coherence is distributed across parameter space. This establishes that coherence is required to generate curvature, but it is the curvature—not the coherence—that controls thermodynamic response.

This connection can be made explicit in the large-dephasing regime. For $\gamma_\phi \gg \gamma,\Omega,\Delta$, the steady-state coherences scale as $x_{\rm ss}, y_{\rm ss} \sim \Gamma_2^{-1}$, implying that the curvature scales as $\mathcal{F}_{\Delta\Omega} \sim \Gamma_2^{-2}$. Consequently, the geometric work vanishes algebraically with increasing dephasing. 

\begin{figure}[t]
\includegraphics[width=\columnwidth]{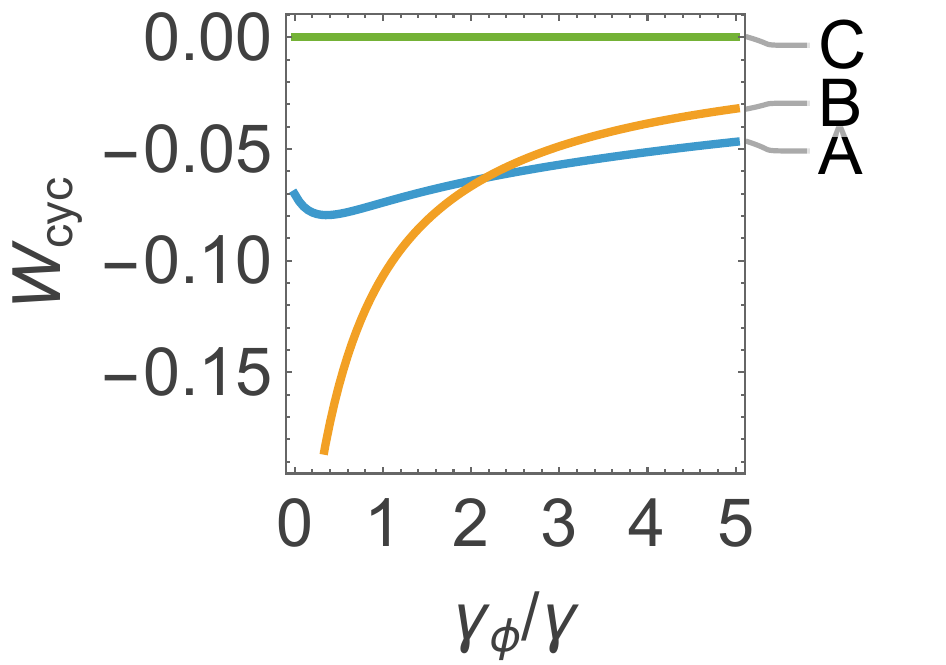}
\caption{
Cycle work $W_{\rm cyc}$ as a function of dephasing for loops A, B, and C. Loop B produces significantly larger work due to its location in a region of larger curvature. Loop C spans regions of opposite-sign curvature and yields $W_{\rm cyc}=0$ due to cancellation of the net curvature flux. All responses decay to zero with increasing dephasing, reflecting the suppression of steady-state coherence and the associated flattening of the curvature landscape.
}
\label{fig2}
\end{figure}

Physically, dephasing suppresses geometric response by flattening the steady-state manifold in parameter space, eliminating the curvature that underlies quasistatic work.
\paragraph{Results} 
We evaluate the quasistatic work $W_{\rm cyc}$ for the three cycles shown in Fig.~\ref{fig1} as a function of dephasing rate $\gamma_\phi$, as shown in Fig.~\ref{fig2}. 
Loop B, which lies in a region of large curvature, produces significantly greater work than loop A, despite both cycles enclosing comparable areas in parameter space. 
By contrast, loop C spans regions of opposite-sign curvature and yields $W_{\rm cyc}=0$, reflecting cancellation of the net curvature flux through the enclosed region. 
Importantly, this vanishing does not indicate the absence of dynamics or energy exchange along the path. Rather, it reflects an exact cancellation of geometric contributions arising from regions of opposite curvature, demonstrating that geometric work is controlled by the signed flux of $\mathcal{F}$ rather than the geometric extent of the cycle.

These results establish that the work is controlled by the curvature of the steady-state manifold rather than the geometric area of the cycle. 
As the dephasing rate increases, all responses decay toward zero, consistent with the suppression of steady-state coherence and the corresponding flattening of the curvature landscape. 
At large $\gamma_\phi$, the distinction between loops A and B diminishes, indicating that dephasing reduces not only the magnitude of the response but also its spatial differentiation in parameter space.

\begin{figure}[t]
\includegraphics[width=\columnwidth]{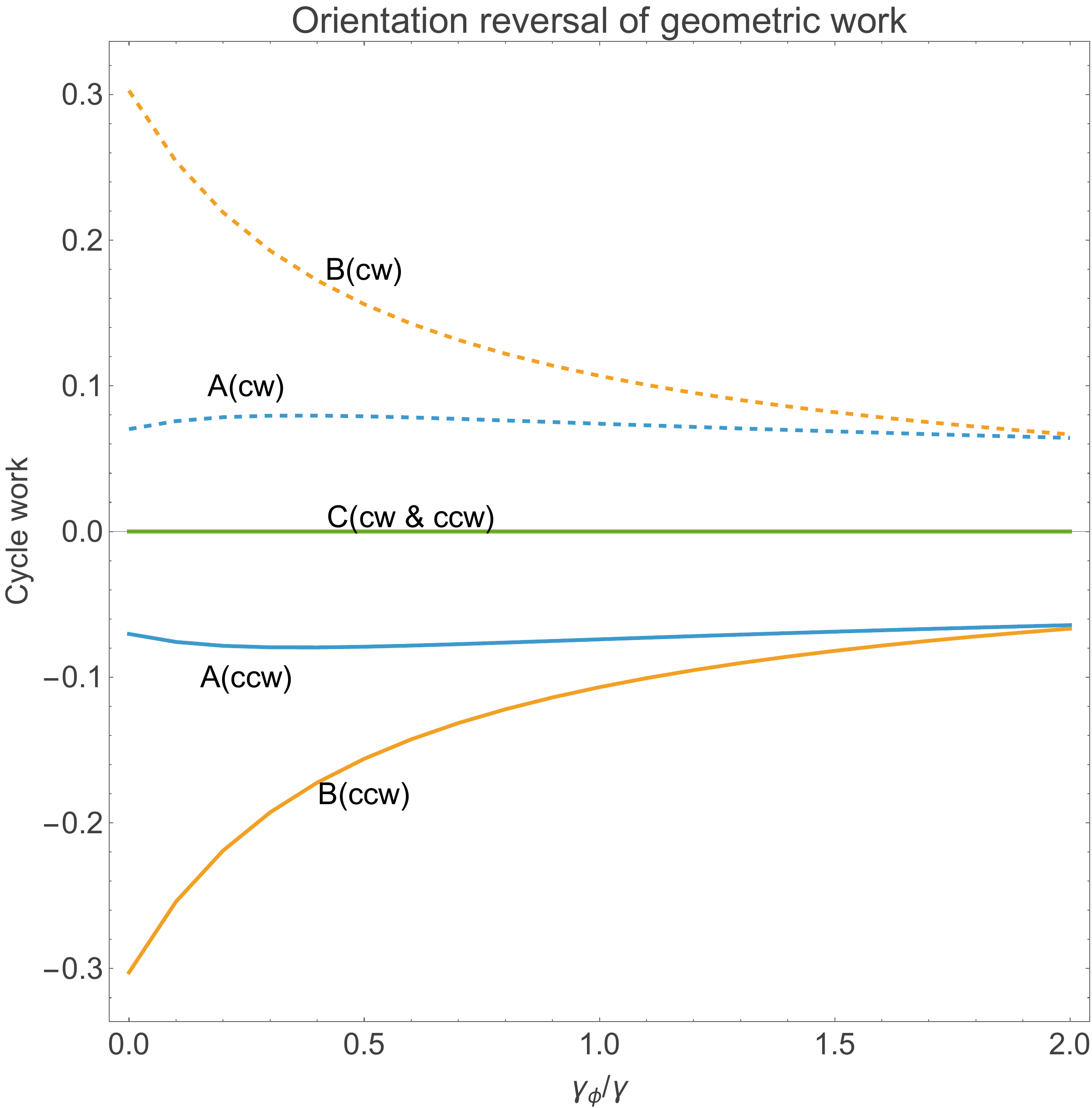}
\caption{
Orientation dependence of the cycle work. Reversing the direction of traversal reverses the sign of the work, confirming its geometric origin.
}
\label{fig4}
\end{figure}

The geometric origin of the work is further confirmed by reversing the orientation of the parameter cycle. As shown in Fig.~\ref{fig4}, the work changes sign under reversal of the traversal direction for both loops A and B, satisfying $W_{\mathcal{C}^{-1}} = - W_{\mathcal{C}}$.
This antisymmetry is the defining signature of a curvature-induced response and establishes that the work arises from a geometric two-form in parameter space. The magnitude of the work decreases with increasing dephasing, consistent with the behavior observed in Fig.~2, while the antisymmetric structure is preserved, confirming the robustness of the geometric description.

The geometric structure defined above admits a natural gauge interpretation. 
The work one-form $\mathcal{A}$ is defined up to the addition of an exact differential,
\begin{equation}
\mathcal{A} \;\rightarrow\; \mathcal{A} + d\chi,
\end{equation}
which leaves the curvature $\mathcal{F} = d\mathcal{A}$ invariant. 
This gauge freedom reflects the fact that only the non-exact (curvature) component of $\mathcal{A}$ contributes to cyclic work, while any exact contribution integrates to zero over closed paths. 
In this sense, $\mathcal{A}$ plays the role of a gauge potential in parameter space, and $\mathcal{F}$ is the associated field strength.

Within this framework, the vanishing of the work for loop C acquires a precise geometric interpretation. 
Although the loop encloses a finite region of parameter space, it samples curvature of opposite sign such that the net flux of $\mathcal{F}$ through the enclosed surface vanishes. 
Equivalently, the line integral of $\mathcal{A}$ along the loop can be decomposed into exact and non-exact components, with only the latter contributing to $W_{\rm cyc}$. 
Loop C therefore isolates a purely exact contribution to the connection. Although the system is driven around a finite path in parameter space, the net work vanishes because only the integrable component of $\mathcal{A}$ is sampled. This demonstrates that geometric work is generated exclusively by the non-exact (curvature) component of the steady-state response.

This decomposition makes explicit the thermodynamic consistency of the geometric response. 
Because only the curvature contributes to cyclic work, and because this contribution is antisymmetric under path reversal, the resulting work is reversible in the quasistatic limit and does not correspond to entropy production. 
The gauge structure therefore separates geometric work—arising from non-integrability—from dissipative irreversibility, which remains encoded in the underlying Lindblad dynamics.

\paragraph{Discussion and Conclusion}
We have shown that open quantum systems with coherent steady states support a new form of thermodynamic geometry. Unlike equilibrium systems, where geometric structure is inherited from an equation of state, the curvature identified here emerges dynamically from the competition between coherent driving and dissipation. This curvature defines a quasistatic work two-form and produces nonzero work in cyclic processes without reference to an equilibrium potential.

Steady-state coherence plays a structural role: it is required for curvature to exist, but does not determine its magnitude. Instead, thermodynamic response is governed by the spatial organization of curvature in parameter space. As a result, cycles enclosing comparable areas can produce markedly different work depending on their location. This establishes a new paradigm in which thermodynamic response is controlled by curvature landscapes rather than scalar state variables.

These results point to a route for engineering thermodynamic response in nonequilibrium quantum systems. In cavity quantum electrodynamics, where coherent control and dissipation can be tuned independently, the curvature of the steady-state manifold becomes a design parameter. More broadly, our work suggests that geometric structure is an emergent organizing principle of open quantum dynamics, with extensions to many-body systems, correlated environments, and symmetry- or topology-protected Liouvillian structures.

An additional mechanism for vanishing geometric work arises from the algebraic structure of the Hamiltonian itself. If the Hamiltonian depends only on a single effective combination of control parameters, the associated work one-form becomes integrable and the curvature vanishes identically. In this case, finite cycles in parameter space yield zero work not through cancellation of positive and negative curvature, but because the underlying response is effectively one-dimensional.

This situation arises, for example, in lattice models at special symmetry points. In the Su--Schrieffer--Heeger (SSH) model, the Bloch Hamiltonian at $k=\pi$ reduces to $H \propto (t_1 - t_2)\sigma_x$, so that the two nominal control directions $(t_1,t_2)$ collapse onto a single effective coordinate in the vacinity of this point. Consequently, the curvature $\mathcal{F}_{t_1 t_2}$ vanishes at $k=\pi$. Cycles symmetric about this point therefore yield zero net work by cancellation, while displaced cycles can enclose regions of finite curvature and generate a nonzero response.

This observation highlights two distinct routes to $W_{\rm cyc}=0$: cancellation of curvature contributions across regions of opposite sign, as in loop C, and algebraic reduction of the control space to an integrable manifold. 
Together, these mechanisms emphasize that geometric work is governed not only by the magnitude of curvature, but by the interplay between parameter-space dimensionality and the algebraic structure of the Hamiltonian.
Exploring how such algebraic structure organizes curvature landscapes in multilevel systems remains an important direction for future work.

\vspace{0.5cm}

\begin{acknowledgments}
This work at the University of Houston was supported by the National Science Foundation under CHE-2404788 and the Robert A. Welch Foundation (E-1337).
\end{acknowledgments}

\paragraph{Data Availability Statement}
All data generated or analyzed during this study are included in this manuscript.

\paragraph{Author Contributions}
The author developed the theoretical framework, performed all derivations,
and carried out the analysis presented in this work. 

\paragraph{Conflicts of Interest}
The author declares no competing financial or non-financial interests.

\bibliographystyle{apsrev4-2}
\bibliography{geom_oqs_literature_v2-3}

\begin{thebibliography}{12}%
\makeatletter
\providecommand \@ifxundefined [1]{%
 \@ifx{#1\undefined}
}%
\providecommand \@ifnum [1]{%
 \ifnum #1\expandafter \@firstoftwo
 \else \expandafter \@secondoftwo
 \fi
}%
\providecommand \@ifx [1]{%
 \ifx #1\expandafter \@firstoftwo
 \else \expandafter \@secondoftwo
 \fi
}%
\providecommand \natexlab [1]{#1}%
\providecommand \enquote  [1]{``#1''}%
\providecommand \bibnamefont  [1]{#1}%
\providecommand \bibfnamefont [1]{#1}%
\providecommand \citenamefont [1]{#1}%
\providecommand \href@noop [0]{\@secondoftwo}%
\providecommand \href [0]{\begingroup \@sanitize@url \@href}%
\providecommand \@href[1]{\@@startlink{#1}\@@href}%
\providecommand \@@href[1]{\endgroup#1\@@endlink}%
\providecommand \@sanitize@url [0]{\catcode `\\12\catcode `\$12\catcode
  `\&12\catcode `\#12\catcode `\^12\catcode `\_12\catcode `\%12\relax}%
\providecommand \@@startlink[1]{}%
\providecommand \@@endlink[0]{}%
\providecommand \url  [0]{\begingroup\@sanitize@url \@url }%
\providecommand \@url [1]{\endgroup\@href {#1}{\urlprefix }}%
\providecommand \urlprefix  [0]{URL }%
\providecommand \Eprint [0]{\href }%
\providecommand \doibase [0]{https://doi.org/}%
\providecommand \selectlanguage [0]{\@gobble}%
\providecommand \bibinfo  [0]{\@secondoftwo}%
\providecommand \bibfield  [0]{\@secondoftwo}%
\providecommand \translation [1]{[#1]}%
\providecommand \BibitemOpen [0]{}%
\providecommand \bibitemStop [0]{}%
\providecommand \bibitemNoStop [0]{.\EOS\space}%
\providecommand \EOS [0]{\spacefactor3000\relax}%
\providecommand \BibitemShut  [1]{\csname bibitem#1\endcsname}%
\let\auto@bib@innerbib\@empty
\bibitem [{\citenamefont {Weinhold}(1975)}]{Weinhold1975}%
  \BibitemOpen
  \bibfield  {author} {\bibinfo {author} {\bibfnamefont {F.}~\bibnamefont
  {Weinhold}},\ }\href@noop {} {\bibfield  {journal} {\bibinfo  {journal} {J.
  Chem. Phys.}\ }\textbf {\bibinfo {volume} {63}},\ \bibinfo {pages} {2479}
  (\bibinfo {year} {1975})}\BibitemShut {NoStop}%
\bibitem [{\citenamefont {Ruppeiner}(1995)}]{Ruppeiner1995}%
  \BibitemOpen
  \bibfield  {author} {\bibinfo {author} {\bibfnamefont {G.}~\bibnamefont
  {Ruppeiner}},\ }\href@noop {} {\bibfield  {journal} {\bibinfo  {journal}
  {Rev. Mod. Phys.}\ }\textbf {\bibinfo {volume} {67}},\ \bibinfo {pages} {605}
  (\bibinfo {year} {1995})}\BibitemShut {NoStop}%
\bibitem [{\citenamefont {Grmela}(2014)}]{Grmela2014}%
  \BibitemOpen
  \bibfield  {author} {\bibinfo {author} {\bibfnamefont {M.}~\bibnamefont
  {Grmela}},\ }\href@noop {} {\bibfield  {journal} {\bibinfo  {journal}
  {Entropy}\ }\textbf {\bibinfo {volume} {16}},\ \bibinfo {pages} {1652}
  (\bibinfo {year} {2014})}\BibitemShut {NoStop}%
\bibitem [{\citenamefont {van~der Schaft}\ and\ \citenamefont
  {Maschke}(2018)}]{vanDerSchaft2018}%
  \BibitemOpen
  \bibfield  {author} {\bibinfo {author} {\bibfnamefont {A.}~\bibnamefont
  {van~der Schaft}}\ and\ \bibinfo {author} {\bibfnamefont {B.}~\bibnamefont
  {Maschke}},\ }\href@noop {} {\bibfield  {journal} {\bibinfo  {journal}
  {Entropy}\ }\textbf {\bibinfo {volume} {20}},\ \bibinfo {pages} {925}
  (\bibinfo {year} {2018})}\BibitemShut {NoStop}%
\bibitem [{\citenamefont {Seifert}(2012)}]{Seifert2012}%
  \BibitemOpen
  \bibfield  {author} {\bibinfo {author} {\bibfnamefont {U.}~\bibnamefont
  {Seifert}},\ }\href@noop {} {\bibfield  {journal} {\bibinfo  {journal} {Rep.
  Prog. Phys.}\ }\textbf {\bibinfo {volume} {75}},\ \bibinfo {pages} {126001}
  (\bibinfo {year} {2012})}\BibitemShut {NoStop}%
\bibitem [{\citenamefont {Lindblad}(1976)}]{Lindblad1976}%
  \BibitemOpen
  \bibfield  {author} {\bibinfo {author} {\bibfnamefont {G.}~\bibnamefont
  {Lindblad}},\ }\href {https://doi.org/10.1007/BF01608499} {\bibfield
  {journal} {\bibinfo  {journal} {Communications in Mathematical Physics}\
  }\textbf {\bibinfo {volume} {48}},\ \bibinfo {pages} {119} (\bibinfo {year}
  {1976})}\BibitemShut {NoStop}%
\bibitem [{\citenamefont {Breuer}\ and\ \citenamefont
  {Petruccione}(2002)}]{BreuerPetruccione}%
  \BibitemOpen
  \bibfield  {author} {\bibinfo {author} {\bibfnamefont {H.-P.}\ \bibnamefont
  {Breuer}}\ and\ \bibinfo {author} {\bibfnamefont {F.}~\bibnamefont
  {Petruccione}},\ }\href@noop {} {\emph {\bibinfo {title} {The Theory of Open
  Quantum Systems}}}\ (\bibinfo  {publisher} {Oxford University Press},\
  \bibinfo {year} {2002})\BibitemShut {NoStop}%
\bibitem [{\citenamefont {Ji}\ \emph {et~al.}(2025)\citenamefont {Ji},
  \citenamefont {Palomino}, \citenamefont {Goldman}, \citenamefont {Ozawa},
  \citenamefont {Riseborough}, \citenamefont {Wang},\ and\ \citenamefont
  {Mera}}]{Ji2025}%
  \BibitemOpen
  \bibfield  {author} {\bibinfo {author} {\bibfnamefont {G.}~\bibnamefont
  {Ji}}, \bibinfo {author} {\bibfnamefont {D.~E.}\ \bibnamefont {Palomino}},
  \bibinfo {author} {\bibfnamefont {N.}~\bibnamefont {Goldman}}, \bibinfo
  {author} {\bibfnamefont {T.}~\bibnamefont {Ozawa}}, \bibinfo {author}
  {\bibfnamefont {P.}~\bibnamefont {Riseborough}}, \bibinfo {author}
  {\bibfnamefont {J.}~\bibnamefont {Wang}},\ and\ \bibinfo {author}
  {\bibfnamefont {B.}~\bibnamefont {Mera}},\ }\href@noop {} {\bibfield
  {journal} {\bibinfo  {journal} {arXiv preprint arXiv:2507.14028}\ } (\bibinfo
  {year} {2025})},\ \Eprint {https://arxiv.org/abs/2507.14028}
  {arXiv:2507.14028 [cond-mat.mes-hall]} \BibitemShut {NoStop}%
\bibitem [{\citenamefont {Carollo}\ \emph {et~al.}(2018)\citenamefont
  {Carollo}, \citenamefont {Spagnolo},\ and\ \citenamefont
  {Valenti}}]{Carollo2018}%
  \BibitemOpen
  \bibfield  {author} {\bibinfo {author} {\bibfnamefont {A.}~\bibnamefont
  {Carollo}}, \bibinfo {author} {\bibfnamefont {B.}~\bibnamefont {Spagnolo}},\
  and\ \bibinfo {author} {\bibfnamefont {D.}~\bibnamefont {Valenti}},\ }\href
  {https://doi.org/10.1038/s41598-018-27362-9} {\bibfield  {journal} {\bibinfo
  {journal} {Sci. Rep.}\ }\textbf {\bibinfo {volume} {8}},\ \bibinfo {pages}
  {9852} (\bibinfo {year} {2018})}\BibitemShut {NoStop}%
\bibitem [{\citenamefont {Crooks}(1999)}]{Crooks1999}%
  \BibitemOpen
  \bibfield  {author} {\bibinfo {author} {\bibfnamefont {G.~E.}\ \bibnamefont
  {Crooks}},\ }\href@noop {} {\bibfield  {journal} {\bibinfo  {journal} {Phys.
  Rev. E}\ }\textbf {\bibinfo {volume} {60}},\ \bibinfo {pages} {2721}
  (\bibinfo {year} {1999})}\BibitemShut {NoStop}%
\bibitem [{\citenamefont {Esposito}\ \emph {et~al.}(2009)\citenamefont
  {Esposito}, \citenamefont {Harbola},\ and\ \citenamefont
  {Mukamel}}]{Esposito2009}%
  \BibitemOpen
  \bibfield  {author} {\bibinfo {author} {\bibfnamefont {M.}~\bibnamefont
  {Esposito}}, \bibinfo {author} {\bibfnamefont {U.}~\bibnamefont {Harbola}},\
  and\ \bibinfo {author} {\bibfnamefont {S.}~\bibnamefont {Mukamel}},\
  }\href@noop {} {\bibfield  {journal} {\bibinfo  {journal} {Rev. Mod. Phys.}\
  }\textbf {\bibinfo {volume} {81}},\ \bibinfo {pages} {1665} (\bibinfo {year}
  {2009})}\BibitemShut {NoStop}%
\bibitem [{\citenamefont {Bittner}(2026)}]{Bittner2026_quantumGeom}%
  \BibitemOpen
  \bibfield  {author} {\bibinfo {author} {\bibfnamefont {E.~R.}\ \bibnamefont
  {Bittner}},\ }\href@noop {} {\bibfield  {journal} {\bibinfo  {journal} {arXiv
  preprint arXiv:2603.22452}\ } (\bibinfo {year} {2026})}\BibitemShut {NoStop}%
\end{thebibliography}%

\end{document}